\documentclass{PoS}

\title{Light-by-light scattering in UPC at the LHC}

\ShortTitle{Light-by-light scattering in UPC at the LHC}

\author{Mariola K{\l}usek-Gawenda\thanks{I would like to thank the DIS2016 Organizing Committee
for a possibility to present our results.}\\
        Institute of Nuclear Physics, Polish Academy of Sciences, Radzikowskiego 152,
PL-31-342 Krak\'ow, Poland\\
        E-mail: \email{mariola.klusek@ifj.edu.pl}}

\author{Piotr Lebiedowicz\\
        Institute of Nuclear Physics, Polish Academy of Sciences, Radzikowskiego 152,
PL-31-342 Krak\'ow, Poland\\
        E-mail: \email{piotr.lebiedowicz@ifj.edu.pl}}

\author{\speaker{Antoni Szczurek}\\
        Institute of Nuclear Physics, Polish Academy of Sciences, Radzikowskiego 152,
PL-31-342 Krak\'ow, Poland and University of Rzesz\'ow, PL-35-959 Rzesz\'ow, Poland\\
        E-mail: \email{antoni.szczurek@ifj.edu.pl}}
\abstract{We discuss diphoton semi(exclusive) production in ultraperipheral $PbPb$ collisions 
at energy of $\sqrt{s_{NN}}=$ 5.5 TeV (LHC). The nuclear calculations are based on equivalent photon approximation 
in the impact parameter space. 
The cross sections for elementary $\gamma \gamma \to \gamma \gamma$ subprocess are calculated including two different mechanisms: box diagrams with leptons and quarks in the loops and a VDM-Regge contribution with virtual 
intermediate hadronic excitations of the photons. We got relatively high cross sections in $PbPb$ collisions ($306$ nb). 
This opens a possibility to study the $\gamma \gamma \to \gamma \gamma$ (quasi)elastic scattering at the LHC. 
We find that the cross section for elastic $\gamma\gamma$ scattering could be measured in the lead-lead collisions
for the diphoton invariant mass $W_{\gamma\gamma} \approx 15-20$ GeV.}

\FullConference{XXIV International Workshop on Deep-Inelastic Scattering and Related Subjects\\
		11-15 April, 2016\\
		DESY Hamburg, Germany}

\begin{document}

\section{Introduction}
In classical Maxwell theory photons/waves/wave packets do not interact.
In contrast, in quantal theory they can interact via quantal fluctuations.
So far only inelastic processes, i.e. production of hadrons or jets
via photon-photon fusion could be measured e.g. in $e^+ e^-$ collisions
or in ultraperipheral collisions (UPC) of heavy-ions.
It was realized only recently that ultraperipheral heavy-ions 
collisions can be also a good place where photon-photon elastic
scattering could be tested experimentally \cite{d'Enterria:2013yra,KGLS2016}.

\section{Theory}
\subsection{Elementary cross section}

\begin{figure}[!h]
\centering
\includegraphics[scale=0.25]{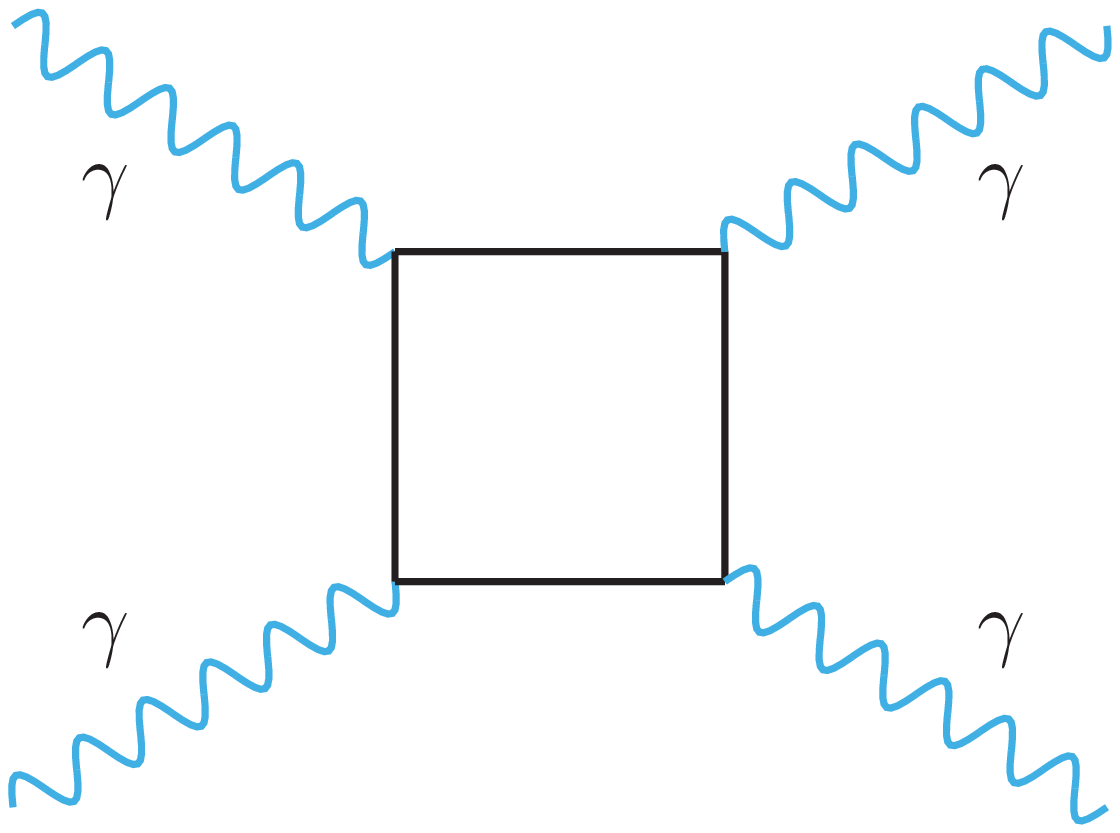}
\includegraphics[scale=0.25]{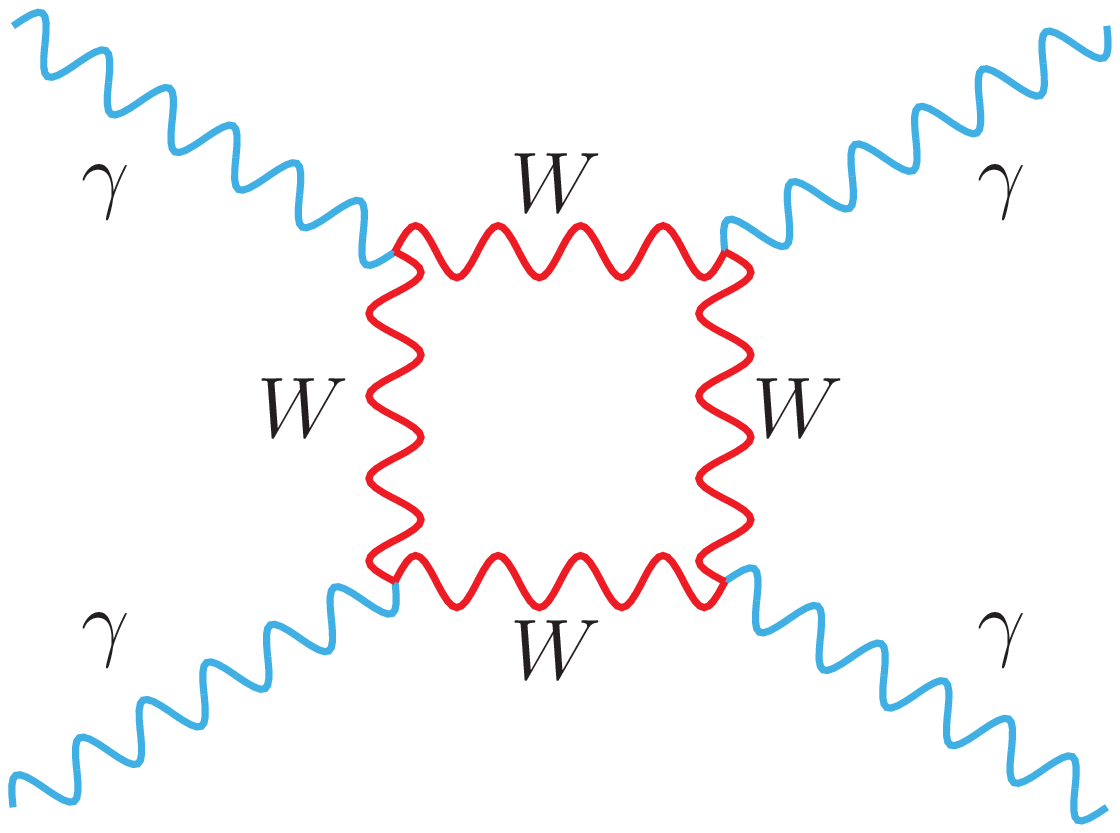}
\includegraphics[scale=0.35]{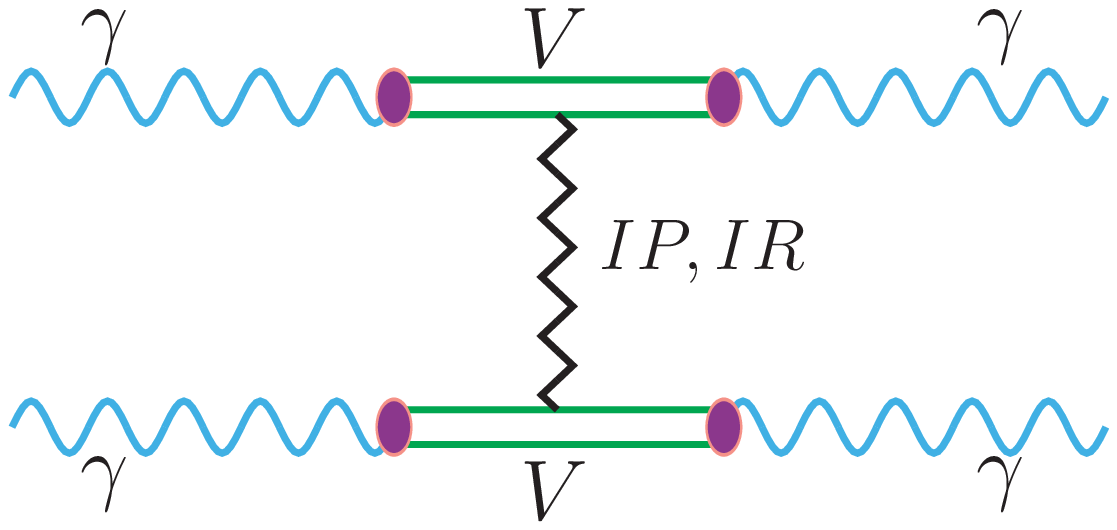}
\includegraphics[scale=0.35]{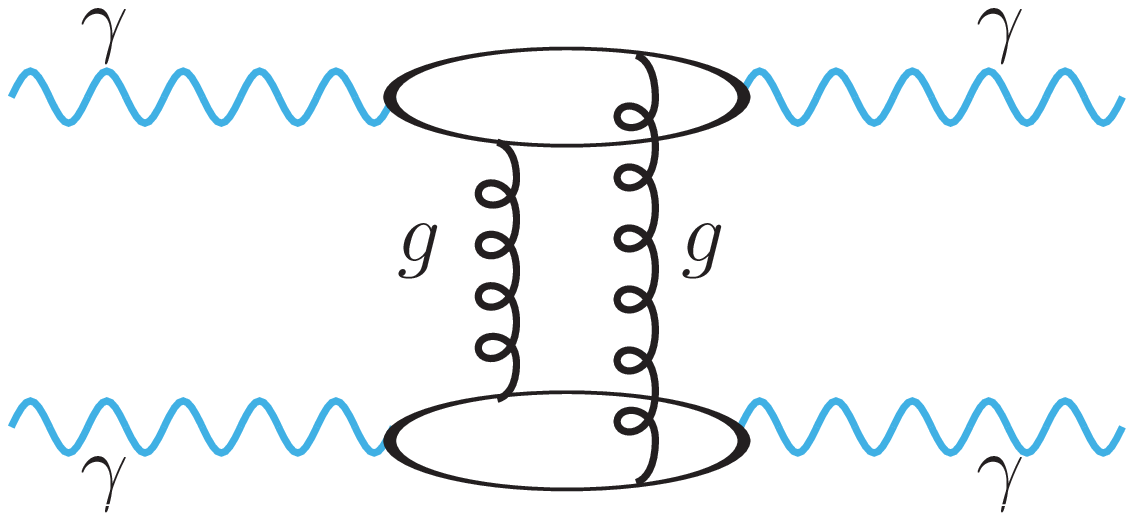}
\caption{Light-by-light scattering mechanisms with 
the lepton and quark loops (first panel) and for the intermediate $W$-boson loop (second panel). 
Third panel represents VDM-Regge mechanism and the last panel is for two-gluon exchange.}
\label{fig:diagrams_elementary}
\end{figure}

One of the main ingredients of the formula for calculation of the nuclear cross section
is elementary $\gamma\gamma\to\gamma\gamma$ cross section. 
The lowest order QED mechanisms with elementary particles are shown
in two first diagrams of Fig.~\ref{fig:diagrams_elementary}. 
The first diagram is for lepton and quark loops and it dominates
at lower photon-photon energies ($W_{\gamma\gamma}<2m_W$) 
while the next diagram is for the $W$ (spin-1) boson loops and it becomes 
dominant at higher photon-photon energies (\cite{Bardin2009,Lebiedowicz2013}). 
The one-loop box amplitudes were calculated by using
the Mathematica package {\tt{FormCalc}} and the {\tt{LoopTools}}
library.
We have obtained good agreement when confronting our result with those in 
\cite{Jikia1993,Bern2001,Bardin2009}.
Including higher-order contributions in the context of elastic scattering 
seems to be interesting.
In Ref.~\cite{Bern2001} the authors considered both 
the QCD and QED corrections (two-loop Feynman diagrams)
to the one-loop fermionic contributions in the ultrarelativistic limit
($\hat{s},|\hat{t}|,|\hat{u}| \gg m_f^2$). 
The corrections are quite small numerically so the leading order computations 
considered by us are satisfactory.
In last two diagrams of Fig.~\ref{fig:diagrams_elementary} we show processes 
that are the same order in $\alpha_{em}$ but higher order in $\alpha_s$. 
In this article we consider process shown in the third panel where
both photons fluctuate into virtual vector mesons
(here we include three different light vector mesons: $\rho, \omega, \phi$). 
The last diagram shows the mechanism which is formally three-loop type but 
we will not consider here the contribution of this mechanism. 
The exact calculations for this process in the context 
of elementary $\gamma\gamma\to\gamma\gamma$ 
and nuclear $AA \to AA\gamma\gamma$ cross section were done in Ref.~\cite{KGSSz2016} .

\subsection{Nuclear cross section}

\begin{figure}[!h]
\centering
\includegraphics[scale=0.25]{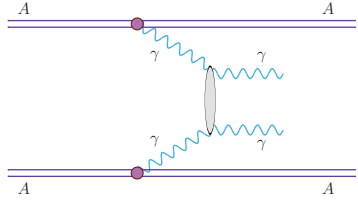}
\caption{Diphoton production in ultrarelativistic UPC of heavy ions.}
\label{fig:diagram_AA_AAgamgam}
\end{figure}
%

The general situation for the $AA \to AA \gamma \gamma$ reaction 
is sketched in Fig.~\ref{fig:diagram_AA_AAgamgam}.
In our equivalent photon approximation (EPA) in the impact parameter space, 
the total (phase space integrated) cross section 
is expressed through the five-fold integral
(for more details see e.g.~\cite{KG2010})
\begin{equation}
\sigma_{A_1 A_2 \to A_1 A_2 \gamma \gamma}\left(\sqrt{s_{A_1A_2}} \right) =\int \sigma_{\gamma \gamma \to \gamma \gamma} 
\left(W_{\gamma\gamma} \right)
N\left(\omega_1, {\bf b_1} \right)
N\left(\omega_2, {\bf b_2} \right) 
S_{abs}^2\left({\bf b}\right)
2\pi b \mathrm{d} b \, \mathrm{d}\overline{b}_x \, \mathrm{d}\overline{b}_y \, 
\frac{W_{\gamma\gamma}}{2}
\mathrm{d} W_{\gamma\gamma} \, \mathrm{d} Y_{\gamma \gamma} \;,
\nonumber 
\label{eq:EPA_sigma_final_5int}
\end{equation}
where $N(\omega_i,{\bf b_i})$ are photon fluxes,
$W_{\gamma\gamma}=\sqrt{4\omega_1\omega_2}$
and $Y_{\gamma \gamma}=\left( y_{\gamma_1} + y_{\gamma_2} \right)/2$ 
is a invariant mass and a rapidity of the outgoing $\gamma \gamma$ system. 
Energy of photons is expressed through $\omega_{1/2} = W_{\gamma\gamma}/2 \exp(\pm Y_{\gamma\gamma})$.
$\bf b_1$ and $\bf b_2$ are impact parameters 
of the photon-photon collision point with respect to parent
nuclei 1 and 2, respectively, 
and ${\bf b} = {\bf b_1} - {\bf b_2}$ is the standard impact parameter 
for the $A_1 A_2$ collision.

The photon flux ($N(\omega,b)$) is expressed through a nuclear charge
form factor. In our calculations we use two different types of the form factor.
The first one, called here realistic form factor, is the Fourier transform of
the charge distribution in the nucleus and the second one is a monopole form factor
which leads to simpler analytical results.
More details can be found find e.g. in \cite{KGLS2016,KG2010}.
 
\section{Results}

\begin{figure}[!h]  
\center
\includegraphics[scale=0.28]{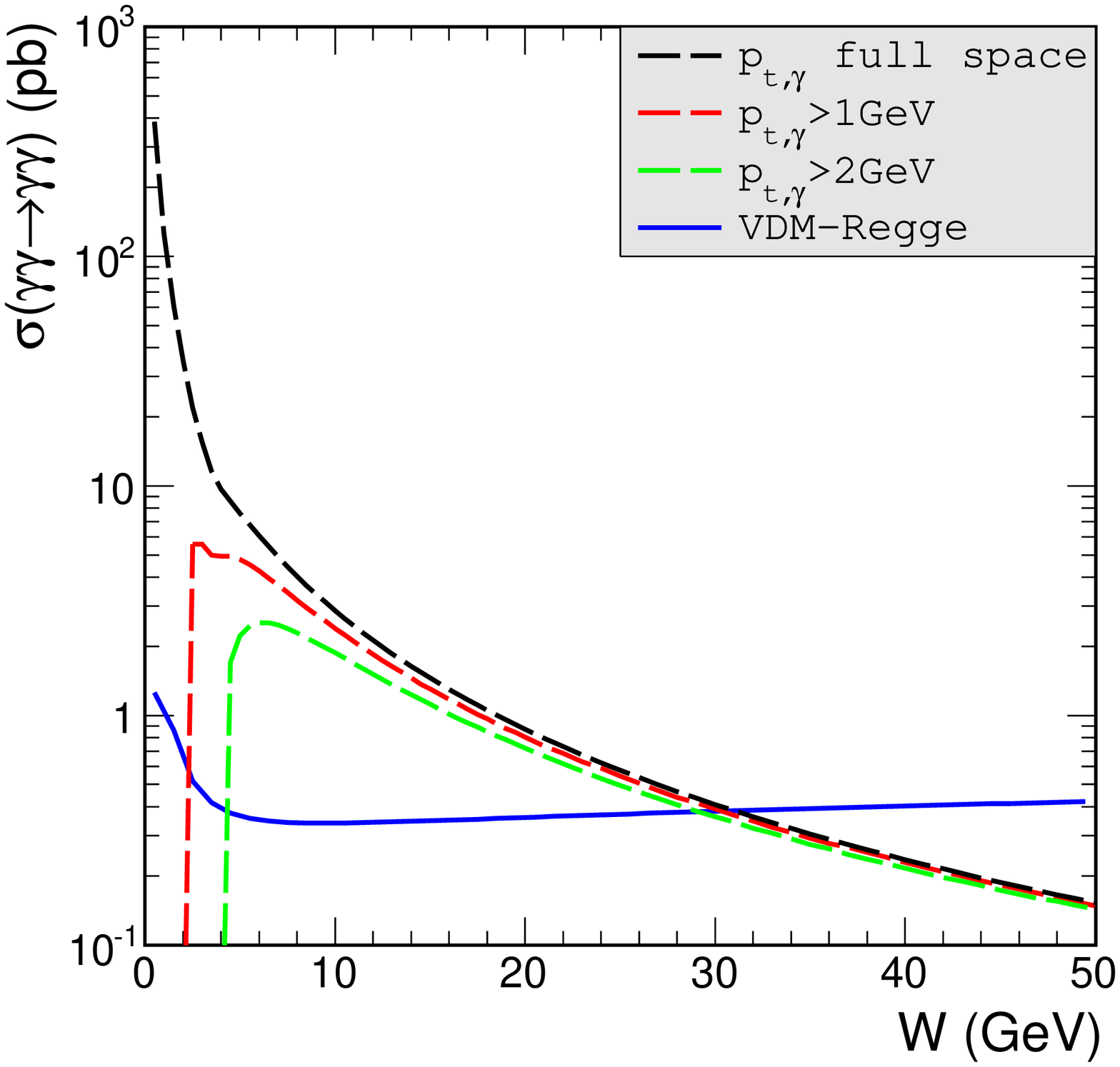}
\includegraphics[scale=0.28]{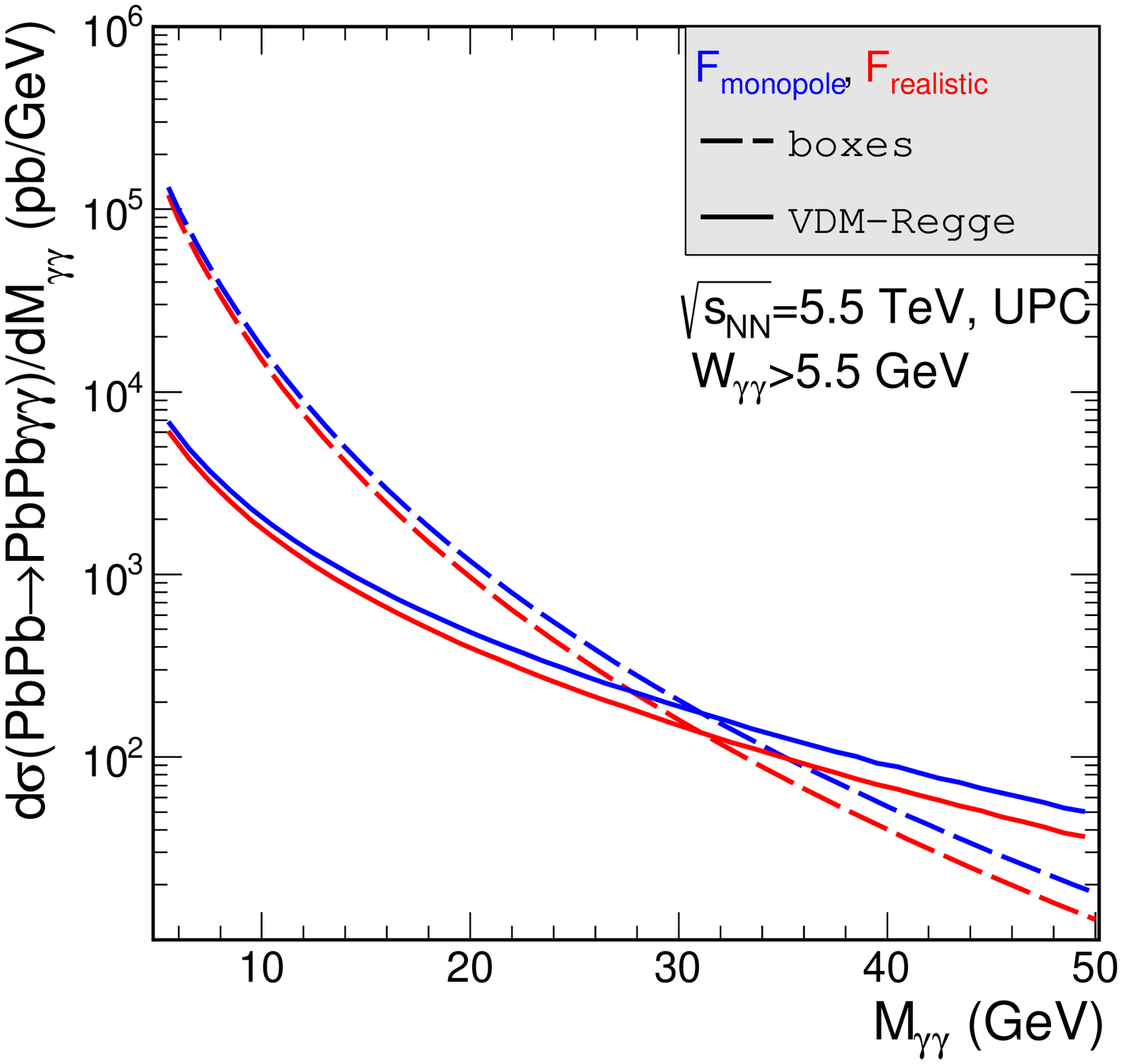}
  \caption{
  Elementary and nuclear cross section for light-by-light scattering.
  The dashed lines show the results for the case
  when only box contributions (fermion loops) are included and 
  the solid lines show the results for the VDM-Regge mechanism.
  Left panel: integrated elementary $\gamma \gamma \to \gamma \gamma$ cross section
  as a function of the subsytem energy. 
  Right panel: differential nuclear cross section as a function of $\gamma\gamma$ invariant mass
  at $\sqrt{s_{NN}}=5.5$ TeV. 
  The distributions with the realistic charge density are depicted by
  the red (lower) lines and the distributions which are calculated
  using the monopole form factor are shown by the blue (upper) lines.
  }
\label{fig:dsig_dw}
\end{figure}

The elementary angle-integrated cross section for the box and 
VDM-Regge contributions is shown in the first panel of Fig.~\ref{fig:dsig_dw}
as a function of the photon-photon subsystem energy.
Lepton and quark amplitudes interfere enhancing the cross section.
For instance in the 4 GeV $<W<$ 50 GeV region, 
neglecting interference effects, the lepton contribution 
to the box cross section is by a factor $5$ bigger than the quark contribution.
Interference effects are large and cannot be neglected.
At energies $W >30$~GeV the VDM-Regge cross section becomes larger
than that for the box diagrams. The right panel of Fig.~\ref{fig:dsig_dw}
shows results for nuclear collisions for the case of realistic charge density (red lines) 
and monopole form factor (blue lines). 
The difference between the results becomes larger with larger values of 
the kinematical variables. The cross section obtained with the monopole 
form factor is somewhat larger.

\begin{figure}
\centering
\includegraphics[scale=0.28]{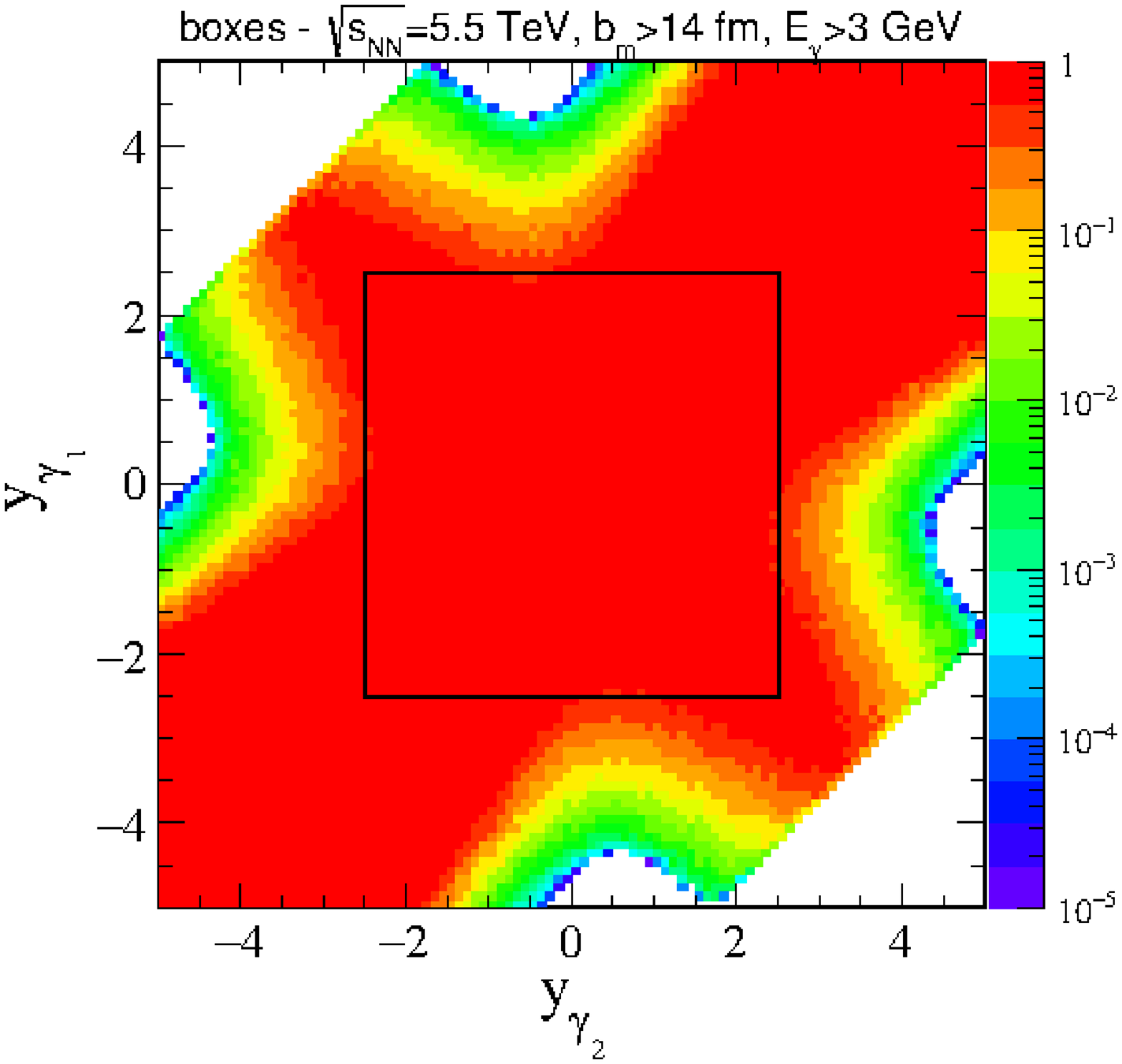}
\includegraphics[scale=0.28]{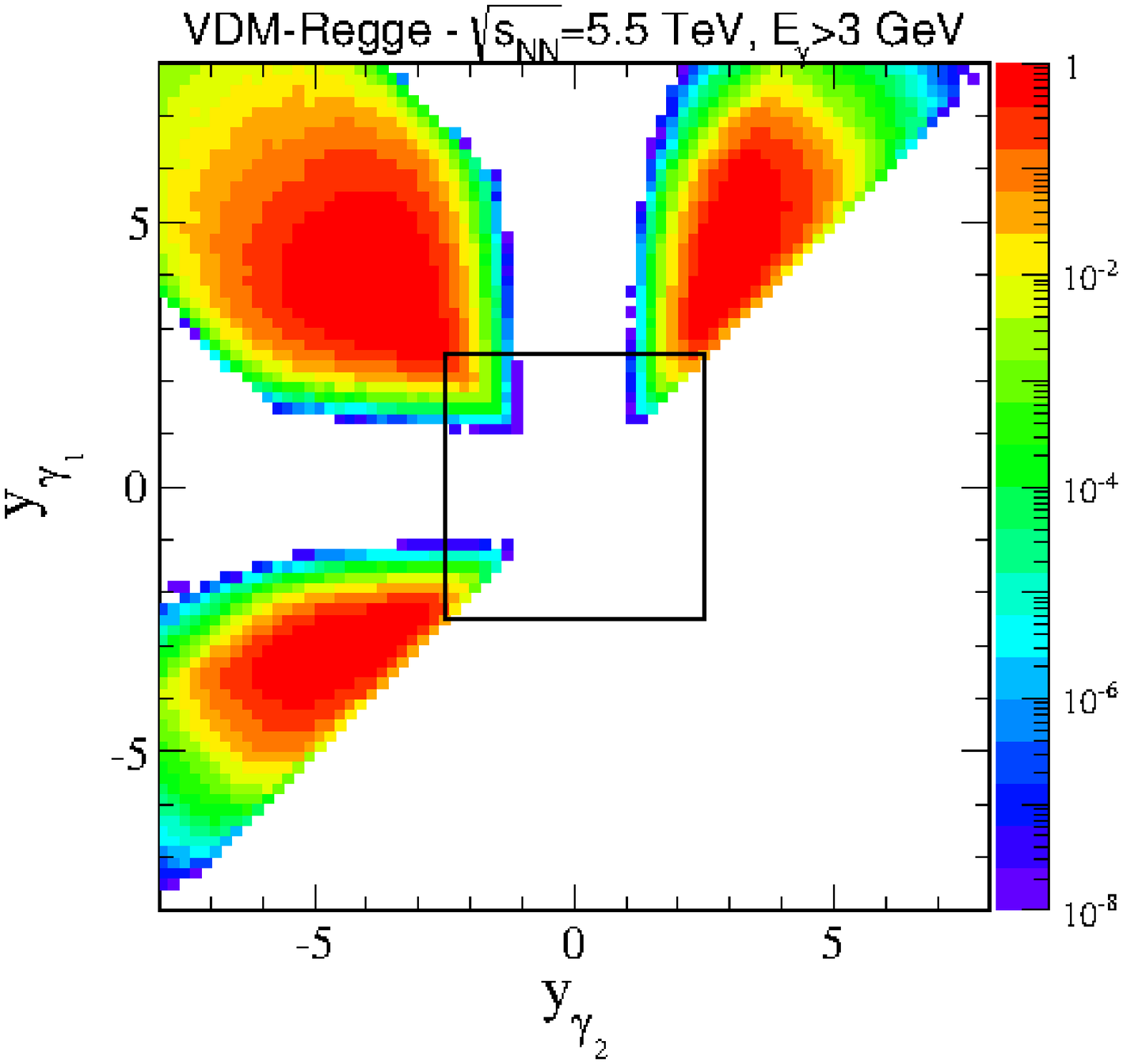}
\caption{\label{fig:dsig_dy1dy2}
Contour representation of two-dimensional 
($\mathrm{d} \sigma / \mathrm{d} y_{\gamma_1} \mathrm{d} y_{\gamma_2}$ in nb) 
distribution in rapidities 
of the two photons in the laboratory frame for box (left panel) 
and VDM-Regge (right panel) contributions.}
\end{figure}

If we try to answer the question whether the reaction can be measured with the help of LHC detectors
then we have to extend Eq.~(\ref{eq:EPA_sigma_final_5int}) by extra integration over
additional parameter related to angular distribution for the subprocess \cite{KGLS2016}.
Fig.~\ref{fig:dsig_dy1dy2} shows two-dimensional distributions in photon rapidities 
in the contour representation. The calculation were done at the LHC energy
$\sqrt{s_{NN}}=5.5$ TeV. Here we impose cuts on energies of photons 
in the laboratory frame ($E_{\gamma}>3$ GeV). Very different distributions
are obtained for boxes (left panel) and VDM-Regge (right panel). 
In both cases the influence of the imposed cuts is significant.
In the case of the VDM-Regge
contribution we observe as if non continues behaviour 
which is caused by the strong transverse momentum dependence of the elementary cross section
(see Fig.~4 in Ref.~\cite{KGLS2016})
which causes that some regions in the two-dimensional space are 
almost not populated. Only one half of the ($y_{\gamma_1},y_{\gamma_2}$) space is shown for
the VDM-Regge contribution. The second half can be obtained from the symmetry
around the $y_{\gamma_1}=y_{\gamma_2}$ diagonal.
Clearly the VDM-Regge contribution does not fit
to the main detector ($-2.5<y_{\gamma_1},y_{\gamma_2}<2.5$) and extends towards large rapidities.
In the case of the VDM-Regge contribution we show much broader range of rapidity
than for the box component. We discover that maxima
of the cross section associated with the VDM-Regge mechanism are at
$|y_{\gamma_1}|,|y_{\gamma_2}| \approx$ 5. Unfortunately this is below the limitations 
of the ZDCs ($|\eta| > 8.3$ for ATLAS (\cite{ATLAS2007})
or $8.5$ for CMS (\cite{Grachov2008})).

\begin{figure}[!h]  
\center
\includegraphics[scale=0.28]{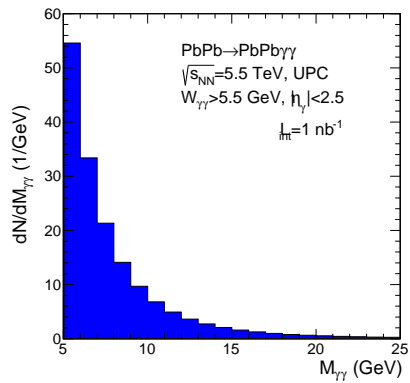}
  \caption{\label{fig:number_of_counts}
  Distribution of expected number of counts in $1$~GeV bins for cuts on 
  $W_{\gamma\gamma}>5.5$ GeV and $\eta_\gamma<2.5$. 
}
\end{figure}

Finally in Fig.~\ref{fig:number_of_counts} we show numbers of counts
in the $1$ GeV intervals expected for assumed integrated luminosity: $L_{int}=1$~nb$^{-1}$
typical for UPC at the LHC.
We have imposed cuts on photon-photon energy and
(pseudo)rapidities of both photons.
It looks that one can measure invariant mass distribution up to 
$M_{\gamma \gamma} \approx 15$ GeV.

\section{Conclusions}

In our recent paper \cite{KGLS2016} we have studied in detail 
how to measure elastic photon-photon scattering
in ultrarelativistic ultraperipheral lead-lead collisions.
The calculations were performed in an equivalent photon approximation 
in the impact parameter space.
The cross section for photon-photon scattering was calculated
taking into account well known box diagrams with elementary standard model particles
(leptons and quarks)
as well as a VDM-Regge component which was considered only recently \cite{KGLS2016}
in the context of $\gamma\gamma \to \gamma\gamma$ scattering.

Several distributions in different kinematical variables were calculated.
Using the monopole form factor we get similar cross section to that found in \cite{d'Enterria:2013yra}
(after the correction given in Erratum of \cite{d'Enterria:2013yra}). 
Nevertheless, we think that application of realistic charge distribution 
in the nucleus gives more precise results.
We have shown an estimate of the counting
rate for expected integrated luminosity. We expect some counts 
for subprocess energies smaller than $W_{\gamma \gamma} \approx$ 15-20 GeV.

\end{document}